**Title:** Micro-cavity length stabilization for fluorescence enhancement using schemes based on higher order spatial modes

**Authors:** A. Shehata Abdelatief,[1,2, a)] A. J. Renders,[1] M. Alqedra,[1,3] J. J. Hansen,[1,4] D. Hunger,[5] L. Rippe,[1] and A. Walther.[1, a)]

[1] Department of Physics, Lund University, P.O. Box 118, SE-22100 Lund, Sweden.
[2] Laser Institute for Research and Applications LIRA, Beni-Suef University, Beni-Suef 62511, Egypt.
[3] KTH Royal Institute of Technology, Roslagstullsbacken 21, 10691 Stockholm, Sweden.
[4] University of Vienna, Faculty of Physics, Vienna Center for Quantum Science and Technology, A-1090 Vienna, Austria.
[5] Physikalisches Institut, Karlsruhe Institute of Technology (KIT), Karlsruhe, Germany.

a) Author to whom correspondence should be addressed: andreas.walther@fysik.lth.se, abdullah.abdelatief@fysik.lth.se

 **Abstract**

We report on experimental investigation of potential high-performance cavity length stabilization using odd-indexed higher-order spatial modes. Schemes based on higher-order modes are particularly useful for micro-cavities that are used for enhanced fluorescence detection of a few emitters, which need to minimize photons leaking from a stabilization beam. We describe the design and construction of an assembly for a microcavity setup with tunable high passive stability. In addition, different types of active stabilization techniques based on higher-order modes, are then implemented and characterized based on their performance. We achieved a stability of about 0.5 pm RMS, while the error photons leaking from the continuous locking beam to a fluorescence detector are suppressed by more than 100-fold. We expect these results to be important for quantum technology implementations of various emitter-cavity setups, where these techniques provide a useful tool to meet the highly challenging demands.

## I.   INTRODUCTION

Optical quantum technologies are a growing field that has attracted the interest of many researchers lately due to the advantages of quantum information processing. In particular, optically detected few-atom systems have recently demonstrated an impressive capacity on a wide variety of platforms,[1-8] targeting applications such as single photon sources, quantum communication, and computation. A key mechanism to increase the photon detection efficiency is the Purcell enhancement of the atomic fluorescence emission enabled by a high-finesse micro/nano optical cavity. Several cavity types have been successful, including those made from a solid block such as ion-beam milled cavities,[9] nano-photonic cavities,[10] WGM disc cavities,[11] but also open access cavities.[12-18] While solid cavity types have the advantage of higher stability against vibrations, a prerequisite for high-finesse operation, they are limited in material selection to what is feasible by production methods. Open access cavities, on the other hand, allow any material that can fit inside the desired microscales, at the cost of being harder to stabilize. The ability to use any material is very important for high performance applications, such as quantum computing, since good coherence properties often requires very special selection of materials, thus motivating the effort of overcoming the stabilization. In Refs. [13-16], many custom designs have been used to improve the mechanical stability based on flexure-mechanical elements such as a nanopositioning stage, passive isolation techniques as well as active feedback. Still, it is challenging to combine stabilization and fluorescence detection, especially when the wavelengths of the stabilization and fluorescence beams are similar, which creates a need for new tools that can help with this trade-off.

In this paper, we investigate different stabilization protocols based on higher-order spatial cavity modes that enable good locking performance while allowing the suppression of error photons leaking through to fluorescence detection by up to several orders of magnitude. Using higher order spatial modes has interesting benefits, e.g., there are several modes available close to the fundamental $TEM_{00}$ mode, such that narrow-band dielectric mirror coatings still allow their use but at wavelengths that can be filtered spectrally compared to the fundamental mode matched to the atoms. In addition, using single mode optical fibers for detecting the fluorescence, either immediately as in many popular fiber cavity approaches[2], or later in the setup like fiber coupled detectors, automatically suppress the higher-order modes used for locking with respect to the fundamental mode. This helps reducing the reliance only on wavelength filters and provides a new tool to meet the challenging needs.

In our experimental demonstrations we also discuss our assembly design, which relies on building all the parts in the same passively damping base housing block, to increase the stiffness and rigidity and reduces the coupling to the disturbance from the environment. We achieve length stability of about 0.5 pm rms at room temperature, while the error photons leaking from the continuous locking beam are measured and remain below the detector's dark counts. We show that error photons from the locking beam can be suppressed by additional two orders of magnitude using higher modes, compared to similar schemes using fundamental orders.

The rest of this paper is structured as follows: Section II describes how coupling part of the light into higher-order modes enables deriving the error signal from a split detector to be utilized in a tilt locking scheme. Section III provides a description of the design and construction of the micro-cavity assembly and its operation. Section IV reports on the experimental methods, locking schemes, higher-order modes coupling to the cavity mode and the optical setup. Section V report on the experimental and theoretical results of mechanical stability measurements using different locking schemes and transfer functions. It also presents the experimental and theoretical results using higher order modes to reduce the error photons to the detector's dark count level.

## II. Higher order modes within cavity

We investigate a fiber-based Fabry-Perot microcavity. The beam that couples light into the cavity can be tilted to selectively excite higher-order Gaussian modes (HOMs), as illustrated in Fig. 1. The spatial transverse electromagnetic (TEM) modes of the cavity can be characterized using Hermite-Gauss functions.[19] For a given cavity geometry, a set of $TEM_{mn}$ modes that meet the round-trip phase condition, where the phase shift must be an integer multiple of $2\pi$, are in resonance with the cavity. In a perfectly aligned and mode-matched cavity, only the fundamental mode, $TEM_{00}$, is resonant with the cavity, while if misalignment between the lock beam and the cavity axis is introduced, higher-order modes (HOMs) resonate in the cavity as well.[20] Particularly, odd-indexed HOMs are used because, in contrast to even HOMs, they have a less effective coupling to the mode of the single-mode fiber (SMF), which, as will be demonstrated later, reduces the amount of stray photons from the locking beam that reach a fluorescence detector. Fig. 1(a) illustrates the intensity distributions corresponding to various cavity $TEM_{mn}$ modes, where n=0 and m takes the values 0, 1, 2, 3, 4, and 5. The mode patterns of the HOMs are spread out considerably farther than the lowest order gaussian mode $TEM_{00}$ by a factor approximately given by $\sqrt{m+1}$.[21, 22] In our cavity, the finite diameter of the fiber mirror results in increased diffraction loss for HOMs, which will be discussed later.

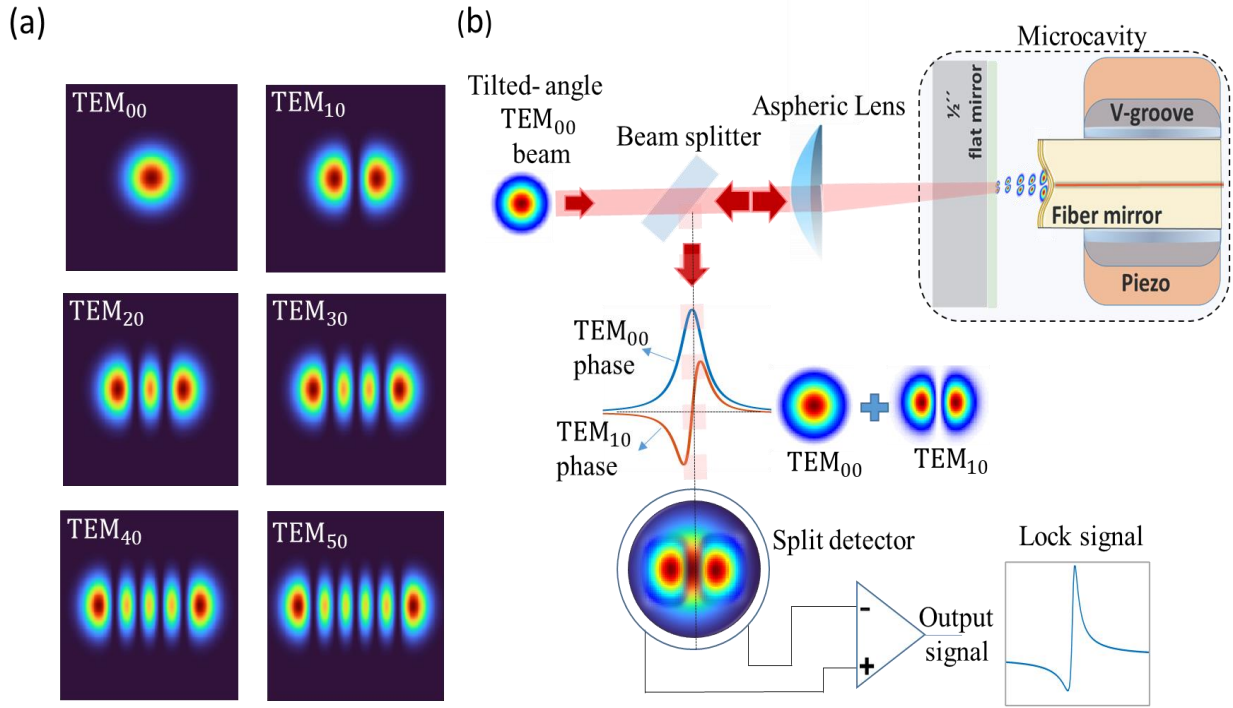

**FIG. 1.** (a) Spatial intensity distribution for the Hermite–Gaussian modes range from $TEM_{00}$ to $TEM_{50}$. The m- and n-modes correspond to the orthogonal directions, and the beam's full transverse profile is obtained from the product of the individual x and y components. (b) The principle of detecting spatial phase gradients in cavity modes to measure cavity length deviations. When there is an angular misalignment in the coupling beam, the $TEM_{10}$ mode couples to the cavity, whereas the symmetric $TEM_{00}$ does not resonate with the same cavity length due to the Gouy phase and is reflected directly. The interference between two modes is detected using a split detector and the difference in the signals from the two sides of the detector can provide insights into the resonance deviations.

The adjacent lobes of produced HOMs (see Fig.1 (a)) have a phase difference of 180°, which can be utilized as a reference to detect deviations in the cavity resonances.[23, 24] In Fig. 1 (b), the input beam is aligned to maximize the coupling of the $TEM_{10}$ mode into the cavity. The $TEM_{10}$ cavity mode detects the deviation from resonance using the $TEM_{00}$ mode as a reference, which is off-resonant and is reflected directly from the flat mirror. The $TEM_{10}$ cavity mode has a relative phase shift of 90° to the $TEM_{00}$ on reflection as seen in Fig. 1(b). The magnitude of the $TEM_{10}$ mode can be quantified by measuring the differential spatial shift resulting from the interference between the field amplitudes of the $TEM_{00}$ and $TEM_{10}$ modes. A segmented photodetector, divided along the y-axis, effectively captures this spatial interference, allowing each lobe of the $TEM_{10}$ mode to be detected separately on each side of the detector. The signals from the two half planes are subtracted, leading to a change in power level which is indicated in Fig. 1(b) as a lock signal. This change reflects the differential constructive and destructive interference between the two $TEM_{10}$ lobes with opposite polarities and the symmetric TEM00 as seen in Fig.1 (b). The output signal can effectively compensate for any changes in the cavity length, and this method is known as a tilt locking.[23, 24]

### III. Device design

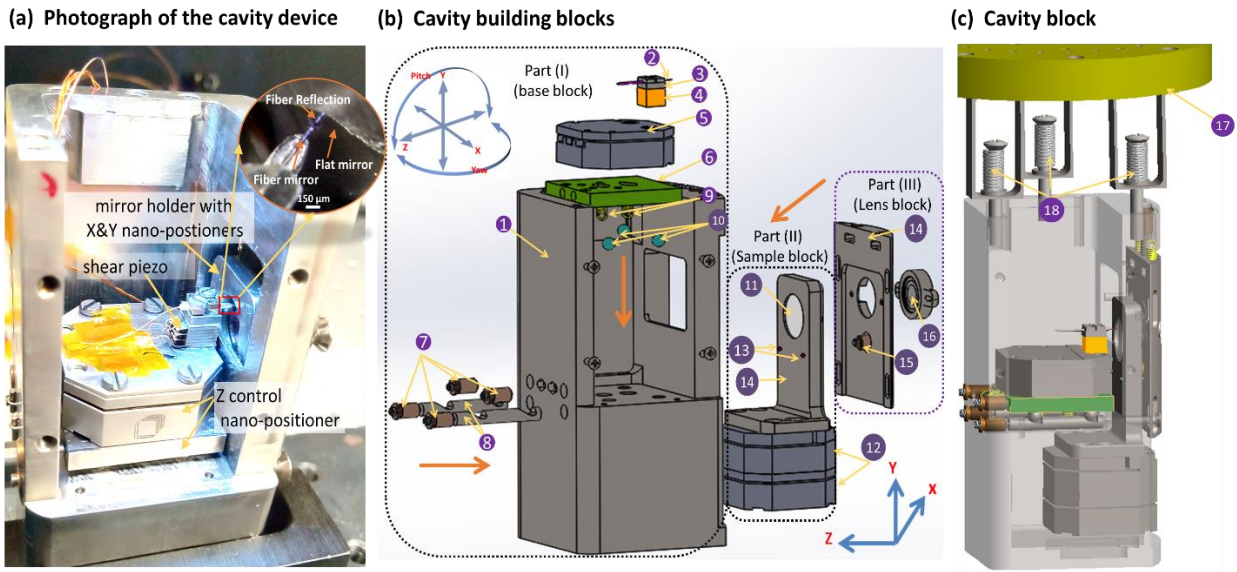

**FIG. 2.** Design and implementation of the micro-cavity assembly (a) a photograph of the constructed cavity assembly and (b) the assembly building blocks, marked with numbers, are described in detail in the text and the orange arrows denote the direction in which the cavity components should be assembled. (c) finished cavity assembly with one transparent side hanging from roof of a cryostat cold plate.

P a g e 4 | 18

Here follows a detailed description of the design of the microcavity setup. The function is described in Section IV. Our design for the open microcavity, as illustrated in Fig. 3 and the inset Fig. 2(a) shows photograph, is a fiber-based concave mirror and flat mirror as a resonator configuration. It also includes fine and coarse tuning translation stages with full access in XYZ directions and two degrees of angular alignment in YZ (Z is the cavity length dimension), as well as a fully tunable aspheric lens for cavity mode matching through the flat mirror. The design relies on building all the parts in the same base housing block in order to increase stiffness and rigidity and reduce the coupling to the disturbance from the environment, thus increasing the passive stability. As shown in Fig. 2(b), the cavity building blocks are divided into three parts: the base block, the sample block, and the lens block. The base block part (I) is the main block where the sample block part (II) and the lens block part (III) are mounted together to the block (1). As depicted in Fig. 2(b) part (I), the fiber mirror (2) is held in a fixed V-groove (3), which is glued on top of a shear piezo (P-121.03T) (4), providing 3 µm tuning range at room temperature (RT), corresponding to six free spectral ranges (FSR), while the range is a factor of 3 lower (two FSR) when operating the cavity at 4 K. The shear piezo is glued to the top part of a piezoelectric nanopositioner (Attocube ANSx150) (5) for high-precision Z-scanning in a scan range of 80 µm (RT). This Z-nanopositioner holding the fiber mirror is then mounted to the top stage (6), which provides four axes of adjustment. The pitch, yaw, Y, and Z positions of this top stage can be adjusted using four fine adjustments, (7) and (8) (Thorlabs F2ES6 Adjuster and F2ESN1P Threaded Bushing), which are employed as actuators. The mounting is performed by using four stretching springs (9) each exerting a force of 10 N, tuned such that the stage slides well on the four stainless steel spheres (10) of 1.5 mm radius. The stage provides 200 um/rev of fine adjustment over 2.5 mm of travel in the Y and Z axes with a pitch and Yaw angles of 0.6°/rev. The top stage (6) has a variety of mounting holes used to fix the top platform and the stage itself to the bottom and rear of the base block (1); reducing the movable degrees of freedom after the right position for cavity operations has been found will increase the passive stability.

    In our case, the cavity is intended to be used for Purcell enhancement of rare-earth ions doped into a nanocrystal that is spin coated on the flat mirror. It is essential that the XY scan range be large enough to ensure that high-quality nanocrystal samples can be found. The sample block (Part II) in Fig. 2(b) consists of a flat mirror (11) that is positioned in the hole in the middle of the mirror holder and two piezoelectric nanopositioners (12) (Attocube ANSz100 and ANSx150) that move on the Y and X axes, respectively. Our utilization is limited at low temperatures, especially for the Y axis, because the nanopositioners travel range on the X axis is 80 µm (RT), 125 µm (LT), and for the Y axis 50 µm (RT), and 30 µm (LT), respectively. Stage (6) can be utilized to give the Y axis sufficient travel range to locate a good scan area containing the desired-sized particles while still at room temperature. At low temperatures, the differential thermal contraction of the nanopositioners and the building block parts is still challenging, in addition to the reduction in the scanning range due to the capacity reduction of the nanopositioners. One can use positioners with longer ranges, such as electric motors and gear such as in Ref [16], or slip-stick type motion but this adds both considerable size and complexity and can contribute significantly to unwanted vibrations. It is necessary to work on minimizing the thermal contraction upon the cooldown of the assembly. Titanium is used for the base block, mirror holder, and translation stage because it is the same material as the nanopositioners, and it has a low thermal expansion coefficient and mechanical stiffness. In addition, two spheres of radius 0.5 mm (13) are glued to the mirror

mount (14) to make physical contact with the base block and pushed against the base block by fine adjuster (15) glued to spring and attached to the lens holder (14) to ensure mechanical stiffness while keeping the flexibility to move the mirror in the XY directions. Spatial mode matching to the microcavity is achieved with an aspheric lens with 0.6 NA (16) glued to a micrometer thread-to-thread adapter. The completed cavity assembly is seen hanging from the cryostat cold plate's roof (17) in Fig. 2(c). The assembly hangs by three springs (18), each with a spring constant of 220 N/m, designed to suppress the resonance frequency of the assembly in addition to the frequencies of about 47 Hz coming from the 1.5 K cell and the compressor that pumps the helium into the cold head.

## IV. Experimental methods and setup

The probe beam is used as a reference beam to measure the cavity length deviations, mostly originating from acoustic and mechanical vibrations, electrical noise, and thermal drifts induced in mirror coatings. For a given cavity mirror coating material ($Ta_2O_5/SiO_2$), the shift in cavity resonance is calculated as 5.5 linewidths per watt of intracavity power,[12] which may affect our experiments only at the highest powers. Relative frequency drift between the lock and probe lasers has to be small. In the case of a lock beam, a single-mode pigtailed distributed feedback (DFB) diode laser (Nanoplus, Germany) operating around 935 nm with an output of about 4 mW is used to stabilize the cavity length. This diode laser is temperature stabilized using a Peltier element and can be both temperature- and current-tuned to ensure that both the lock beam and the probe beam at 892 nm can simultaneously match their respective cavity resonances. The temperature drifts and current drifts from the controllers induce a frequency drift ranging from 21 MHz to 56 MHz over 24 hours. This is sufficient for good lock beam stability during the measurements. In the case of the probe beam, a stabilized CW Ti: sapphire laser at 892 nm with a linewidth of about 50 kHz, and about 2% fluctuation in power is observed. The lock beam is coupled to the cavity mode from the flat mirror side by using a mode-matching aspheric lens to gain the advantage of the angular alignment to the cavity mode.

Two locking schemes are tested and compared for stabilizing the cavity length and measuring the performance of the locking setup: the side-of-fringe locking and tilt locking schemes. In the case of side-of- fringe locking, the cavity length is tuned to be resonant to one of the reflected HOMs and the probe beam simultaneously, and the cavity length is actively stabilized at half of the reflection dip of the lock beam at 935 nm and the length deviations is monitored by the probe beam. As discussed in Section II, the tilt locking scheme uses the phase shift induced by resonant light as the cavity drifts out of resonance.[23, 24] For the cavity with a few microns of mirror separations, HOMs are resonant for slightly different lengths on the order of nm due to the additional Gouy phase shift. In this study, the first-order mode is used to stabilize the cavity length by tilt locking to achieve the benefit of low coupling to the single mode fiber. When the first-order mode is coupled to the cavity, it is affected by the phase shift, while the symmetric zero-order mode is not resonant to that cavity length and is directly reflected. Similar to the sidebands in the Pound-Drever-Hall (PDH) technique, [25, 26] the symmetric part of the reflected light is used as a reference signal in this case. The interference between the reflected zero-order mode and the resonant first-order mode will be a spatially asymmetric shape and can be measured by a vertical split detector. An error signal with a near-linear phase response can then be generated by taking the difference of the detector's two vertical halves.

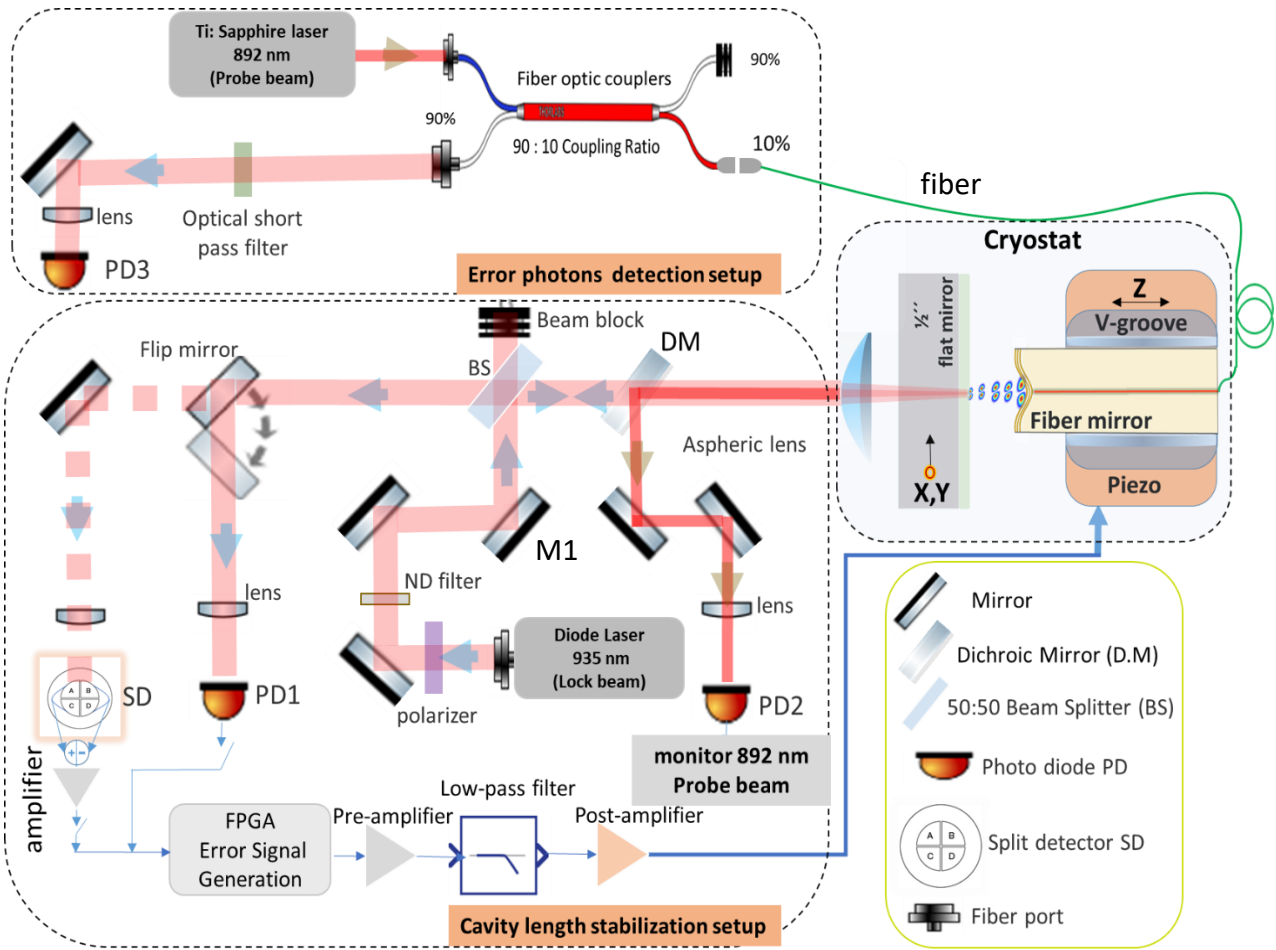

**FIG. 3.** Overview of relevant optical and electrical components in the experimental setup. Different optical schemes used for measuring the cavity length stabilization using higher order modes (HOMs), and detection of error photons in a fluorescence setup leaking through the fiber cavity. Detection methods are described in the main text.

Figure 3 shows a schematic of the electronic and optical setup used to stabilize the cavity length, which is clamped to an optical table resting on air dampers. The microcavity used in our setup consists of an end-facet single-mode fiber mirror machined using a CO2 laser ablation technique,[12] and a ½'' flat mirror. The coatings of the microcavity mirrors are designed to have $10^3$ ppm mirror transmission at an 892 nm probe beam and about $2 \times 10^3$ ppm at a 935 nm lock beam. At 892 nm, a finesse of 3000 is measured, and at 935 nm, we observed a finesse of 1600 at the fundamental mode. The microcavity is operated in reflection and transmission, as shown in Fig. 3. In cavity transmission, the two wavelengths at 892 nm and 935 nm are separated by a dichroic mirror (DM). A collimated diode laser beam at 935 nm is sent to the mode-matching lens of the open-cavity device, which couples light to the cavity. The reflected light travels back along the same path and is redirected by a 50% transmission beam splitter. Using a flip mirror, one can select between the vertical split detector, which is used in the tilt locking method for cavity length stabilization, and the PD1, which is used in the side-of-fringe locking scheme. In the side of fringe locking analysis, the transmitted light from the beam splitter is reflected by the flip mirror and focused on a photodiode PD1 (Hamamatsu, S5973) connected to a transimpedance FEMTO amplifier (DHPCA-100) set to a gain of $10^7$ V/A and a bandwidth of 220 kHz. For the tilt locking analysis, a silicon quadrant photodiode (First Sensor QP5.6-TO5) is placed into a rotatable holder to adjust the splitting axis of the detector and mounted on a micrometer stage that is adjustable along the XYZ axes of the incoming beam. To generate the error signal used in the cavity length stabilization, the detector is used in a vertically split mode by electrically connecting the right and left quadrants, and the subtraction is done by using a low-noise preamplifier (SRS, Model SR 560). A variable neutral density filter (ND) is held in front of the lock beam to vary the laser power entering the cavity. The arbitrary rotation of the polarization in the fiber mirror, due to its ellipticity,[27] is corrected by using a polarizer after the diode laser output. This helps to suppress unwanted splitting in the excited higher-order modes. A 10% Ti: sapphire laser probe beam at 892 nm is coupled from the fiber side to the cavity through one of the fiber optic coupler's input ports. The transmitted light from the cavity is reflected by a dichroic mirror and focused on another photodiode, PD2, with similar settings as PD1. Using the other fiber coupler port with 90% transmission, the error photons transmitted from the fiber when the cavity is stabilized at various HOMs are measured on PD3. Although not included in this paper, when the setup is used for atomic fluorescence measurements, those would also be collected from this output. When the cavity length is stabilized to half of the reflection dip as in the side of fringe locking or near-linear phase response as in the tilt locking scheme, a small deviation in the cavity length resonance is translated to an error signal. This error signal is then sent through a combination of a "proportional-integral" (PI) controller, a low-noise pre-amplifier (SRS, Model SR 560), a first-order RC low-pass filter, and a post-amplifier (TD250) before finally being applied to the shear piezo (P-121.03T).

## V. Results

### A. Coupling of higher order modes

The scheme shown in Fig. 3 is used to perform the cavity length stabilization measurements. Again, the cavity is locked to the diode laser at 935 nm coupled to the cavity mode from the flat mirror side, and the error signal is monitored by Ti: sapphire laser at 892 nm coupled to the cavity from the fiber side. The stability measurements are performed with different HOMs coupled to the cavity with a series of input power levels ranging from 50 nW to 4 mW for each. The probe beam is coupled to the cavity as a fundamental mode during the experiments, and its input power is fixed at $P_i$ = 260 nW. Assuming unity fiber coupling and mode matching efficiency, for a mirror transmission T = $10^3$ ppm and finesse F = 3000, this power corresponds to the intracavity power of $P_c = P_i T (F/\pi)^2$ = 237 µW. From calculations, this is far below the level that may cause photothermal effects due to the probe beam intracavity power,[12, 28] as we shall see in Sec V.B. In order to excite the HOMs inside the cavity, as shown in Fig. 4 (a), a well-aligned lock beam is first coupled to the fundamental mode of the cavity from the flat mirror side by an aspheric lens. Then, by tilting one of the coupling mirrors M1, a misalignment is introduced between the lock beam and cavity axis, resulting in higher-order mode coupling. Each HOM's coupling efficiency mainly depends on the cavity length, incident tilt angle, and off-axis offset, in addition to the mode-matching aspheric lens position, which needs to be precisely aligned. To explore the tilt angle dependence of the HOMs we note that the coupling efficiency scales with $\alpha(\theta/\theta_D)^n$, where θ is the tilt angle, $\theta_D$ is the divergence angle, and n is the HOMs order. Fig. 4 (a) shows a typical cavity reflection signal on the detector PD1 and the transmission signal from the SMF mirror on PD3 when the cavity length is scanned across one FSR. This figure is obtained by blocking the probe beam at 892 nm and slightly tilting the lock beam with the M1 mirror to gain insight into the spectrum of different HOMs excited inside the cavity, and to observe the coupling of the HOMs to the single mode fiber displayed on a logarithmic scale to accentuate the differences in transmission. As shown in Fig. 4 (a), up to 5 HOMs are excited in one free spectral range (FSR) with different reflection dips due to the different coupling efficiencies of HOMs to the cavity mode. Note that the transmission from the fiber is measured when the coupled beam is not optimized for a specific HOM, meaning the figure mostly shows the frequency locations and overall structure. When the resonance of the different HOMs is subsequently optimized, the highest coupling efficiencies observed for the different HOMs are $\eta_{00} \approx 0.65$, $\eta_{10} \approx 0.6$, $\eta_{30} \approx 0.26$ and $\eta_{50} \approx 0.17$, respectively.

For the 935 nm lock beam, we observed an optimized finesse of 1600 for $TEM_{00}$ and $TEM_{10}$, 800 for $TEM_{30}$, and 400 for $TEM_{50}$. Because of the finite size of the fiber-based mirror, the Hermite-Gaussian (HG) modes, in particular the higher order ones, may get altered in shape due to the clipping loss from the mirror.[22] This means that a later experiment with more optimal mirrors would allow higher finesses, which is encouraging for employing the HOMs for cavity length stabilization while the error photons, which come from the transmission, are strongly reduced. Fig. 4(b) gives a typical example of losses introduced to various HOMs coupled separately to the cavity mode with a mode radius $w_m$ impinging on a fiber mirror of different sizes. This figure is obtained by calculating the eigenvalues of the cavity after one round trip, taking into account the finite size of the fiber mirror (see supplementary material 1). The fiber mirror used in this measurement, with a measured 31 μm radius of curvature (ROC) and 1.65 μm depth, gives a mirror radius of about 10.5 μm.[12] As shown in Fig. 4(b), a fiber mirror with a 10.5 μm radius achieves performance not limited by clipping loss for the fundamental mode $TEM_{00}$ and the first-order mode $TEM_{10}$, explaining the measured finesse of 1600 for $TEM_{00}$ and $TEM_{10}$. But in the $TEM_{30}$ and $TEM_{50}$ mode cases, Fig. 4(b) shows the $TEM_{30}$ and $TEM_{50}$ modes get altered significantly with high values of clipping loss of about $3 \times 10^3$ ppm and $1 \times 10^4$ ppm, respectively. Let us consider that the coating of the mirror is the same for HOMs and only clipping loss is introduced to the modes. The finesse corresponding to the calculated clipping loss is about 900 for $TEM_{30}$ and 450 for $TEM_{50}$ which is close to the measured finesse. To achieve performance not limited by the diffraction from the mirror fiber, a fiber mirror with a radius $a > 3 w_m$ is needed to couple up to 5 odd indexed HOMs to the cavity without additional clipping loss, which also agrees with the reported value.[22] Fig. 4(c) shows an example of cavity reflection when the first-order $TEM_{10}$ mode is coupled to the cavity from the flat mirror side together with the transmission of the probe beam. The half-linewidth point of the transmitted probe beam at 892 nm is used for the probe measurements, and is marked in the figure. The inset in Fig. 4(c) shows the time trace signal monitored on PD1 (the reflection lock beam) and on PD2 (the transmission probe beam) before subtracting the locking point. Fig. 4(d) shows the difference error signal from the split detector with the characteristic steep slope, which can be used to stabilize the cavity length. This figure is obtained when the cavity length is scanned over the first-order resonance of the lock beam. The light reflected by the cavity has two peaks with a phase difference of 180°; therefore, one will constructively interfere with the symmetric zero-order mode reflected light while the other destructively interferes. The cavity is tuned to have an overlap between the probe beam and the steep slope difference error signal for cavity length stabilization. To investigate the cavity length fluctuations, the transmission time trace of the probe beam is recorded for 10 seconds, and a fast Fourier transform (FFT) is used to obtain an amplitude spectral density (ASD) in units of cavity length fluctuations. In the figure, five such ASDs were recorded and the average is presented.

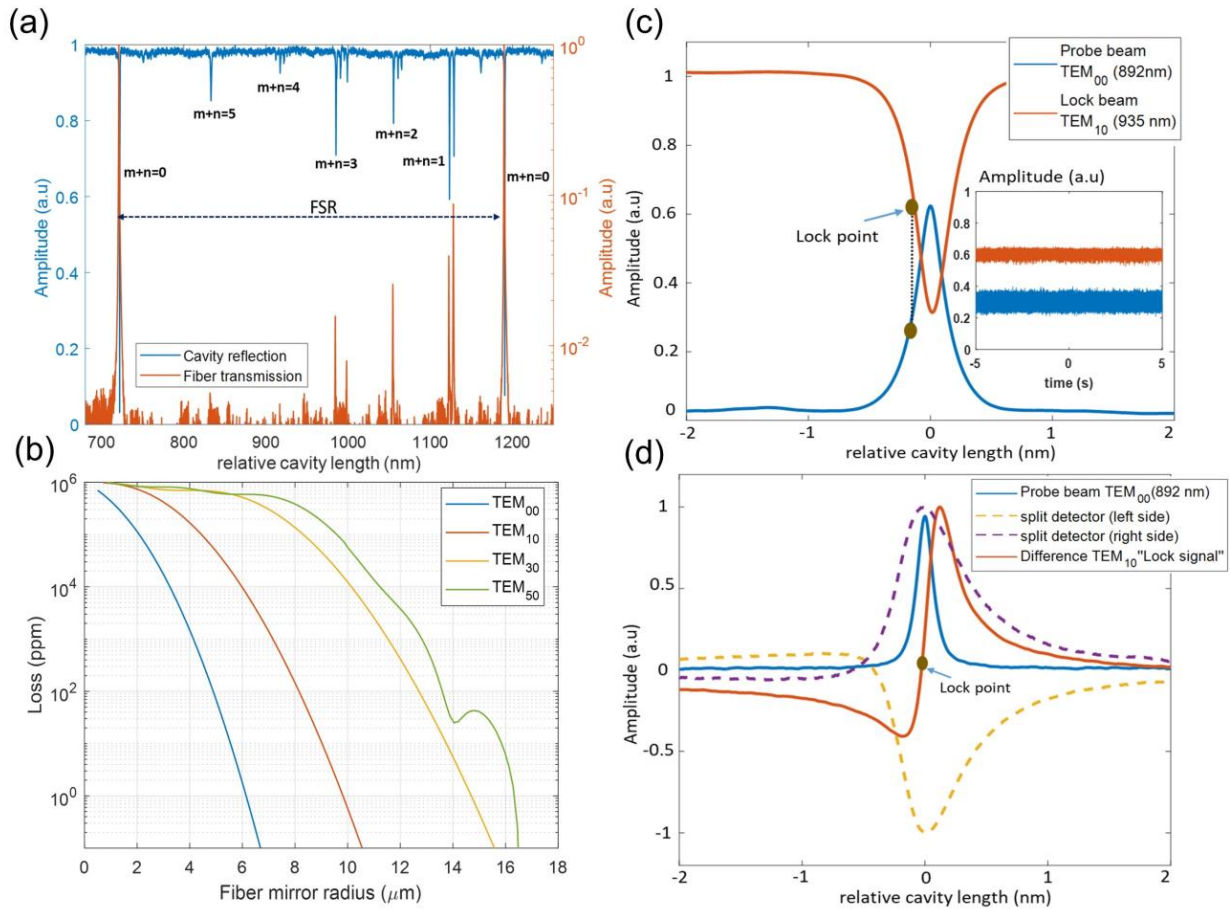

FIG. 4. (a) cavity reflection spectrum from the flat mirror side (blue) and lock beam transmission from the fiber mirror (orange), see Fig. 2. (b) simulation of the clipping loss introduced from the fiber mirror with different finite sizes for different higher order modes excited in a cavity. This clipping loss affects both the finesse of the higher order modes and the coupling between the cavity and fiber modes. (c) the overlap between the transmitted probe beam and the reflected lock beam is shown together with the locking point. The inset is the time trace signal from both signal lock and probe before subtracting the locking point using side of fringe locking. (d) shows the components of the split detector used to tilt locking scheme (orange) and the difference used as a lock signal, and the probe beam (blue).

## B. Performance of microcavity length stabilization

In Fig. 5, we show data measured by the side of fringe locking and tilt locking schemes. ΔL is the relative position change between the two mirrors integrated over all frequencies and given in units of pm rms. The closed-loop bandwidth of the setup is 3 kHz, which is limited by the bandwidth of the post-amplifier used to amplify the signal applied to the shear piezo. Fig. 5(a) shows the ASD for different HOMs when the cavity is stabilized using the side of fringe and tilt locking schemes at the same intracavity power of about 32 mW, which is closer to the thermal limit than the probe beam. The cumulative cavity length fluctuations are utilized to identify the frequencies with the most significant contribution, as illustrated in Fig. 5(b). There are mechanical resonances that appear at 261 Hz and up to 10 kHz. The primary source of mechanical noise is frequencies below 1 kHz, specifically 261, 315, and 470 Hz. The origin of resonance frequencies is identified through multiple measurements (see supplementary material 2 for more information). The first three resonances at 261Hz, 315Hz, and 470 Hz are acoustic noises coupled from the optical table to the cavity, identified through measurements with an independent accelerometer, in addition there are some electronic noises from the PI controller close to the peaks of 261 Hz and 315 Hz (RedPitaya STEMlab 122-16). The next resonance frequencies that appear at 770 Hz, 1.3 kHz, and 3.9 kHz are the assembly's vibration response identified through a resonance frequency analysis using SolidWorks. In addition, the resonance at 3.9 kHz matches the electronic noise from Z-nanopositioners. The resonance at 10.5 kHz is likely acoustic noise, as it falls within the frequency band range (10 Hz-10 kHz) measured by the accelerometer. The side of fringe locking scheme experiences slightly increased rms length fluctuation with higher order modes going from 1.3 pm (TEM$_{00}$) to 2.5 pm (TEM$_{50}$), as depicted in Fig. 5(b). This is due to the lower finesse of the higher orders, as explored in more details later. In a similar fashion, the higher slope of the tilt locking scheme (can be seen by comparing the slopes of Figs. 5(c) and 5(d) respectively) gives an improved stability compared to fringe locking. In total, it gives a four times improvement in stability at the same intracavity power and the same HOM, TEM$_{10}$.

As illustrated in Fig. 5(c), the length fluctuation integrated over all frequencies is studied as a function of the intracavity power for the locking beam in order to obtain insight into the constraints of using different HOMS and locking schemes. To gain further understanding of the limitations, we have modelled our system by a transfer function response. For this, every component of our locked circuit is expressed as a complex transfer function, which allows us to calculate the noise coupling to the cavity and approximate the response as displacement deviations in frequency (see supplementary material 3). Fig. 5(d) shows the results from the transfer function estimations. The error is scaled by a term proportional to $1/(1 + C_s \mathrm{H_{PI}} \mathrm{H_A})$, where $C_s$ is the transfer function from the fiber mirror position to the transimpedance detector's output, $\mathrm{H_{PI}}$ is the transfer function of the PI controller, and $\mathrm{H_A}$ is the transfer function of the preamplifier with low pass filter. The intracavity power increases the term $C_s$, and when the magnitude of $C_s \mathrm{H_{PI}} \mathrm{H_A}$ is large, the influence of displacement perturbations is reduced and the error tends toward its lowest value, which is constrained by electronic noise and the feedback loop's bandwidth (BW). The measured and computed length deviations in Figs. 5(c) and 5(d) show more stability (less deviations) for $TEM_{00}$ and $TEM_{10}$ in comparison to $TEM_{30}$ and $TEM_{50}$. This is because the feedback gain depends on the slope of the error signal, which itself depends on the finesse and contrast of the reflected signal near resonance. Higher-order modes have lower finesse and are thus able to achieve less feedback gain without also increasing noise. More specifically, for an optimized feedback loop the properties of the electronics in the feedback loop, such as transimpedance gain and PI parameters, depend on the optical characteristics of the cavity, such as the finesse and the coupling efficiency of different modes. The finesse and coupling efficiencies of the locking beam are mode-order dependent, as was previously described. Consequently, in the case where $TEM_{00}$ and $TEM_{10}$ are coupled to the cavity, the optical gain is higher and can thus better suppress the error with lower noise. In order to achieve matching between the measured length deviations in Fig. 5(c) and the computed in Fig. 5(d), the proportional gain of the PI controller is increased by a factor of 3.5 and 2.8 to compensate for the lower finesse of 400 and 800, respectively.

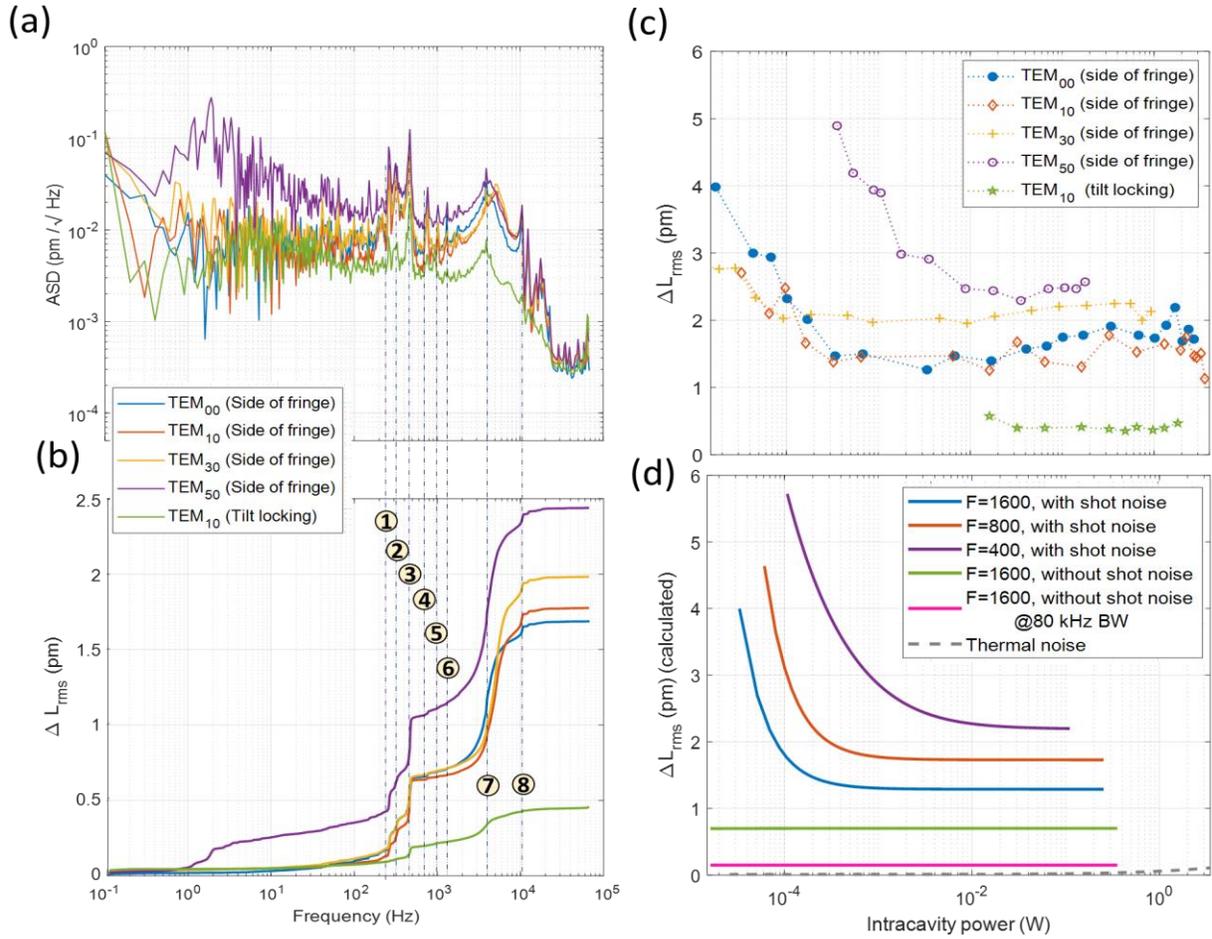

Fig. 5. cavity length stabilization measurements. (a) FFT of the time trace recorded for 10 s at different higher-order modes with intracavity power of 32 mW. (b) cumulative length fluctuation up to a frequency of 60 kHz at different HOMs. The resonance frequencies corresponding to 1 - 8 are 261 Hz, 315Hz, 470 Hz, 771 Hz, and 1 kHz, 1.3 kHz 3.9 kHz and 10.5 kHz, respectively, and their origin is discussed in the main text. (c) length fluctuation integrated over all frequencies as a function of the intracavity power for the locking beam, for each of the locking schemes and orders. (d) Shows the effect on the length fluctuation from locking finesse, shot noise of the beam, closed loop bandwidth, Brownian noise and photothermal noise.

As shown in Figs. 5(c) and 5(d), the shot noise of the detector causes the length deviations for the different modes coupled to the cavity to increase at low intracavity power. With no light incident, the transimpedance detector PD1 used for the side of fringe locking scheme exhibits an integrated output peak-to-peak detector noise of 36 mV at $10^8$ V/A transimpedance gain. For the photodiode detector with 0.26 A/W responsivity and 220 kHz bandwidth, the power $P_0$ that would be necessary for the shot noise of the beam to exceed the detector noise is 7 µW. This power $P_0$ is equivalent to 4 mW intracavity power at 1600 finesse. As shown in Figs. 5(c) and 5(d), for an incident power less than 7 µW, the photocurrent level exceeds the detector noise. For the quadrant split detector used for tilt locking error signal generation, the minimum detectable power is 12 pW per element. However, the noise from the reverse-biased operating circuit and the operational amplifier used to read out the two sides of the split detector and subtract them, respectively, increases this minimum detectable power. Along with the shot noise constraint, there are a number of other considerations that can affect the achievable level of length fluctuation suppression. As shown in Fig. 5(d), these include the closed-loop bandwidth and thermal noises with different origins, such as the Brownian noise and photothermal noise.[12, 28, 29] Photothermal noise is produced by absorbed optical power in mirror coatings and substrate, and it is estimated to be about 113 fm rms for a mirror with $10^3$ ppm absorption at 1 W intracavity power. This value decreases with intracavity power and at some point, approach the Brownian noise limit, which is about 15 fm rms at 32 mW intracavity power. Assuming a perfect coupling to the cavity, if shot noise is ignored, the lowest length fluctuations that can be obtained with a 3 kHz bandwidth feedback loop is 0.5 pm (See Sec. 3 of the supplementary material). This may be reduced further to 150 fm when the bandwidth is increased to 80 kHz. The bandwidth is determined by the entire electro-mechanical response (electronics + amp + piezo + mechanical load); at the same time, the electronic noise from the nanopositioners and the post-amplifier limit the suppression to give fluctuations of about 0.5 pm rms, which is reached by our measured tilt locking scheme.

### C. Locking beam error photons

Improving the performance of the single photon detection coupled to the cavity is a promising route toward scalable quantum information processing and computing. In the previous discussion, an attempt to improve the cavity length stabilization performance was investigated by using different HOMs and different locking schemes. In this section, we work on improving the performance of the single-photon detection by reducing the error photons from the locking beam leaking through the single mode fiber (SMF) during the measurements. Fig. 6 shows the detected and computed error photons coupled to the fiber from the odd-indexed HOM locking beam at various intracavity powers and cavity lengths.

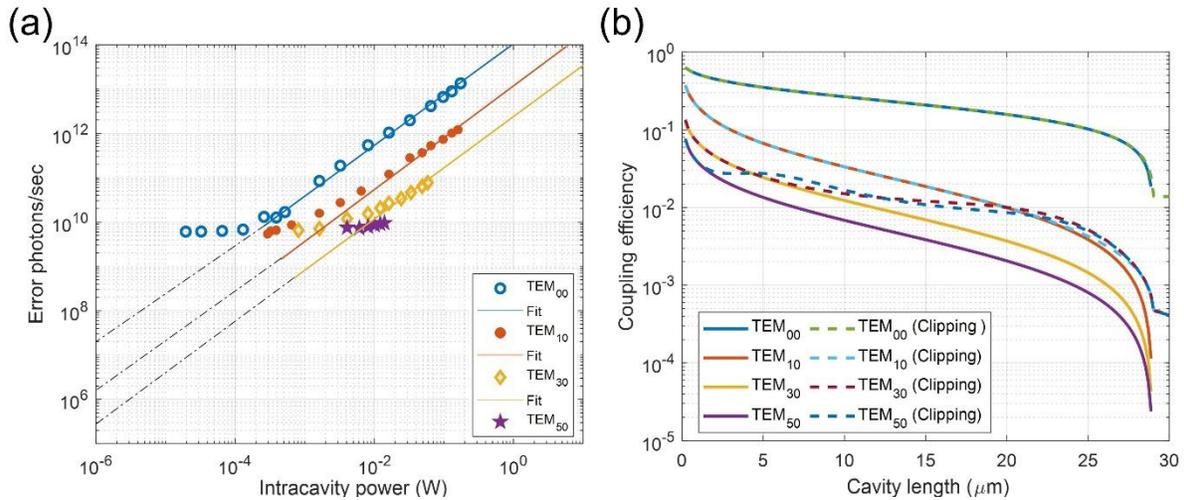

Fig.6. shows the error photons leaking through the single mode fiber mirror when the cavity is stabilized to different HOMs. (a) the error photon rate measured by PD3 (see Fig 3) at different HOMs with varying power for each mode together with the fit line to extract the coupling efficiencies of different modes to the fiber. The lower plateau represents the photodiode detection limit, though the lines show that when using locking powers of $10^{-4}$ W, still allowing good stabilization, using higher order modes will suppress the error photons by about two orders of magnitude. (b) shows the coupling efficiency between the cavity mode and the single mode fiber, both when the size of the mirror is included and without. Thus, even further suppression of error photons could be obtained if larger mirrors were used that then would allow even higher orders.

The error photon rate shown in Fig. 6(a) is obtained by the error photon detection setup (see Fig. 3) while the probe beam is blocked, thus showing the signal from the locking beam itself. Fig. 6(a) is measured by the side of fringe locking scheme when the cavity length is stabilized to half the reflection dip of different HOMs coupled to the cavity with varying input powers. The transmitted light from the fiber is measured by the transimpedance photodetector PD3, which gives an output signal in volts. The measured amplitude in voltage is converted into a photon rate at 935 nm. As shown in Fig. 6 (a), the fundamental mode of $TEM_{00}$ has a high error photon rate compared to the other higher-order modes because the overlap between the cavity and the fiber becomes successively lower for higher order modes, as Fig. 6(b) illustrates. When the intracavity power decreases, the error photons reach the detector's noise limit and are no longer picked up by the regular detector. In order to verify the scaling behavior, a single photon detector is used in place of the PD3, and measurements are obtained using a short-pass filter with an optical density of around $10^4$. Even though the locking is not stable at such low powers, at a 5 µW intracavity power, a single photon detector measures 365 and 77 photons per second for $TEM_{10}$ and $TEM_{30}$, respectively. The detector's 20 photons per second dark count is subtracted from the overall error photon rate to determine the actual error photons from the lock beam. There is no obvious change in the measurement of error photons between the $TEM_{30}$ and $TEM_{50}$ modes, as shown in Fig. 6(a). This can be explained by clipping loss from the finite-size fiber mirror. Fig. 6(b) shows that, except for the zero and first-order modes, the clipping loss significantly contributes to the coupling of the HOMs to the single mode fiber when the finite size of the fiber mirror is taken into account (see supplementary material 4). A fit can be made to the data in Fig. 6(a), where the photon rate transmitted from the fiber mirror through the cavity to the detector is fitted by $J_{out} = P_c \eta \, (T/E)$. Here, $P_c$ is the intracavity power, E=2.12 x10$^{-19}$ J, is the energy per photon and $\eta$ is the total probability to detect photons. Using a known cavity mirror transmittance T of 2 x10$^3$ ppm at a wavelength of 935 nm. The mode coupling to the SMF is a component of the total probability $\eta$ by $\eta = \eta_c \, \eta_{SMF} \, \eta_{setup} \, \eta_{det}$, where $\eta_c$ is the measured mode coupling to the cavity, $\eta_{setup}$ =0.22 is the measured setup efficiency and $\eta_{det}$=0.26, is the detector quantum efficiency. The fitting yields the values of $\eta_{SMF}$ for TEM$_{00}$, TEM$_{10}$ and TEM$_{30}$ of 0.24, 0.03, and 0.012 for, respectively. This agrees with Fig. 6(b)'s estimates at cavity length 10 µm.

## VI. Conclusion and outlook

In conclusion, we presented the layout and operation of a custom-designed, fully-tunable open-access microcavity that could potentially be used for quantum optical technologies and cavity QED investigations. The platform provides full access tuning in XYZ directions with a pitch and Yaw angles better than 100 µrad, as well as additional tuning capability of the aspheric lens for an efficient coupling of light from free space into the cavity mode. We tested the design's performance with different higher-order modes as well as two different locking schemes. The finesse dependence of the mode number was measured, and found to arise from clipping loss introduced by the finite size of the fiber mirror. We achieve mechanical stability of better than 0.5 pm rms using a tilt locking scheme that had the best performance. Using the fringe locking scheme, we measured various mechanical stabilities dependent on the mode excited in the cavity, which ranged from 1.3 pm rms to 2.5 pm rms. Note that the stability for the higher order modes was limited by mirror clipping in our case, and can thus be avoided in future setups. The limitations of the closed loop bandwidth, in addition to the electronic noise of the piezo drivers, determine our final stabilization capability to 0.5 pm. Through our analysis, we have thus identified ways to improve our setup, and suggest how similar setups could be optimally constructed. For instance, it can be technically improved by using high-voltage amplifiers that can operate with a wide bandwidth and low noise level. For example, using an amplifier, such as the ADHV4702-1 with a noise level of around 8 nV/√Hz and a bandwidth higher than the feedback loop, the electronic noise from the nanopositioners and the post-amplifier can be reduced by a factor of 50 for each while not being limited by the bandwidth. Furthermore, implementing the photothermal self-stabilization technique can lower the rms noise amplitude by about one order of magnitude and, in the optimal case, improve the total lock bandwidth to about 500 kHz.[28] According to transfer function estimates, combining such large bandwidth with low noise can result in mechanical stability of roughly 20 fm rms, which is close to the thermal noise limit of 15 fm rms. We have also demonstrated that the error photons leaking from the continuous locking beam can be suppressed by more than two orders of magnitude by using locking schemes based on higher order modes. Extrapolating to future setups with higher finesse cavities, our results predict that a cavity finesse of at least 2x10$^7$ can be used with the locking performance demonstrated here. This shows that open access micro cavity designs are highly competitive platforms even for challenging quantum technologies based on single atom interactions.

**Supplementary Material**

In this supplement, we present the calculations outlined in the main text.

**Acknowledgement**

This research was supported by the European Union FETFLAG program Grant No. 820391 (SQUARE), the Swedish Research Council (2021-03755), and The Royal Physiographic Society in Lund."

**AUTHOR DECLARATIONS**

**Conflict of Interest**

The authors have no conflicts to disclose.

**Author Contributions**

A. Shehata Abdelatief: Conceptualization; Data curation; Formal analysis; Experimental Investigation; Methodology; Writing – original draft; Writing – review & editing. A. J. Renders: Experimental Investigation; Data curation (supporting); Formal analysis (supporting), review & editing. M. Alqedra: Conceptualization, review & editing. J. J. Hansen: Conceptualization, review & editing. D. Hunger: Methodology, review & editing. L. Rippe: Formal analysis (supporting), Methodology (supporting); review & editing. A. Walther: Conceptualization, Project Supervision (lead); Resources (lead); Writing – review & editing (lead).

DATA AVAILABILITY

The data that support the findings of this study are available from the corresponding author upon reasonable request.